# PERFORMANCE ENHANCEMENT OF MEDIUM TEMPERATURE BAKED NIOBIUM SRF CAVITY BY SURFACE CONTAMINATION REMOVAL*

V. Chouhan[†], D. Bice, A. Cravatta, A. Murthy, A. Netepenko, T. Ring, D. Smith, G. Wu
Fermi National Accelerator Laboratory, Batavia, Illinois, USA

## Abstract

Medium-temperature (mid-T) baking, typically conducted at 300–350°C, enhances the quality factor of niobium (Nb) superconducting radio frequency cavities. High vacuum furnace baking is commonly preferred for its practicality in large-scale processing. However, surface contamination, such as niobium carbide formed during vacuum furnace baking, can limit the quench field and degrade the quality factor of the cavity. To investigate this effect, a 1.3 GHz single-cell Nb cavity underwent mid-T baking, followed by a chemical treatment to remove the surface contaminants. Post-treatment measurements revealed a significant improvement in both the quality factor and the quench field.

## INTRODUCTION

Superconducting radio frequency (SRF) cavities made of high RRR niobium are key components in modern particle accelerators, enabling efficient acceleration of charged particles through their high gradients ($E_{acc}$) and quality factors ($Q_0$). To further reduce RF losses and improve operational efficiency, various surface treatment techniques have been developed, including nitrogen doping [1] and thermal baking. Medium-temperature (mid-T) baking in the range of 300–350°C has emerged as a promising method to enhance the $Q_0$ of niobium cavities, particularly at medium RF fields [2, 3]. This technique alters the near-surface composition by oxygen diffusion, suppressing BCS surface resistance and RF losses [2]. Mid-T baking at 350°C has been selected for the PIP-II low-beta 650 MHz cavities [4]. Mid-T baking can be performed either in situ, with the cavity actively pumped during the baking, or by placing the cavity in a high-vacuum furnace. The latter method is generally preferred for mass production of cavities.

Despite the observed $Q_0$ enhancement, mid-T baked cavities often quench at relatively low accelerating gradients of below 25 MV/m. The origin of this reduced quench field remains under investigation. One hypothesis is that niobium carbides, potentially formed during mid-T baking in the furnace, contribute to degraded cavity performance.

To investigate this possibility, a post-bake surface treatment that removes only the topmost RF-active layer without significantly altering the oxygen concentration is desirable. In this work, we explore the application of an ultralight chemical removal process, removing the top RF layer, to a mid-T baked cavity. The goal is to evaluate the effect of surface contaminants on cavity performance and determine whether this minimal surface treatment can improve the performance of mid-T baked cavities.

## CAVITY SURFACE PROCESSING

A 1.3 GHz Nb single-cell cavity (TE1RI010) was selected for the mid-T baking study. The cavity was initially processed with bulk EP, degassed at 900°C for 3 h, and baked at 120°C for 48 h to establish its baseline performance. The cavity was then treated with 5 μm EP to eliminate 120 °C bake effect before adding mid-T baking at 350°C for 3 hours. Following this, ultralight EP, designed to remove approximately 100 nm of the top RF layer, was applied to evaluate the impact of any surface contaminants formed during the mid-T bake. The cavity was tested in a vertical cryostat after each major processing step. Figure 1 shows the sequence of processing and testing steps applied to the cavity.

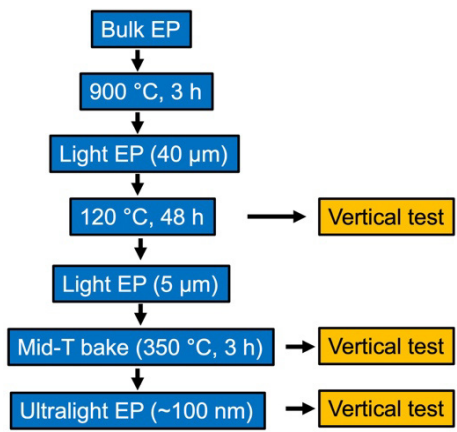

Figure 1: Processing steps applied to the 1.3 GHz single-cell cavity.

### Mid-T Baking

Mid-T baking of this cavity was performed in a vacuum furnace, usually used for degassing of the cavity. The cavity flanges were covered with Nb caps to protect the cavity's inner surface from direct line-of-sight exposure to the furnace heaters. The temperature was ramped up at a rate of 3 °C/min until it reached 350°C, where it was held constant for 3 h to complete the bake, followed by furnace cooling. Figure 2 presents the temperature and pressure profiles recorded during the mid-T baking process.

### Ultralight EP

An ultralight EP was applied to the mid-T baked surface with the aim of removing the top 100 nm surface. This was intended to eliminate any contaminants formed within the RF surface during the mid-T bake process, while preserving both the bulk oxygen concentration and the macroscopic surface morphology. To precisely control the ultralight removal in the EP process, the temperature at the cavity equator was maintained at ~6°C, which was significantly lower than that used in standard cold EP, to reduce the removal rate. EP was carried out for an average removal thickness of 108 nm over a period of two minutes, which allowed the cavity to complete two rotations on the horizontal EP tool. This ensured uniform material removal around the entire cavity circumference. The current density profile and average removal thickness during the ultralight EP process are shown in Fig. 3.

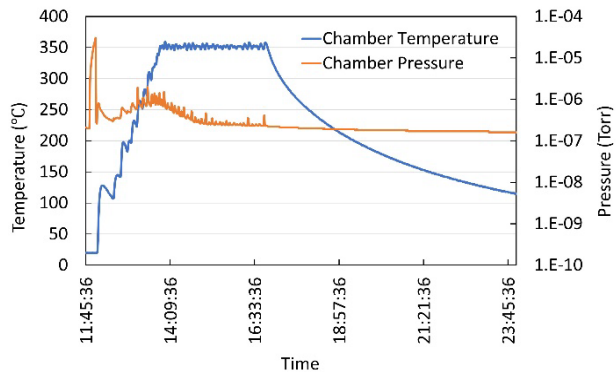

Figure 2: Temperature and pressure profiles during the mid-T bake process.

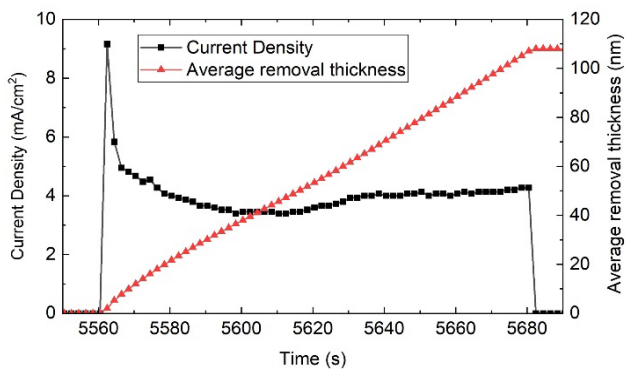

Figure 3: Current density and average removal thickness during ultralight EP performed for two minutes.

## VERTICAL TEST RESULTS

The cavity was tested in a vertical cryostat at 2 K and lower temperatures to evaluate and compare its performance after each processing step, including the 120 °C bake, the mid-T bake, and the post-mid-T ultralight EP. $Q_o$ versus $E_{acc}$ curves from the three tests are shown in Fig. 4.

Following the 120°C bake, the cavity quenched at an accelerating gradient of 47 MV/m corresponding to a peak magnetic field $B_p$ of ~200 mT. After the mid-T bake, the quench field dropped significantly to 22 MV/m ($B_p$ = 93.5 mT), although $Q_o$ improved compared to the baseline.

After applying ultralight EP, the $Q_o$ value improved significantly with a pronounced anti-Q slope. Moreover, the accelerating gradient increased from 22 MV/m ($B_p$ = 93.5 mT) to 32 MV/m ($B_p$ = 137 mT). Notably, no field emission was observed in any of the three tests. The highest $Q_o$ value for mid-T baked surface was observed at a medium field of ~17 MV/m. A comparison of $Q_o$ values at 17 MV/m and quench fields in all three tests is provided in Table 1. At 32 MV/m, $Q_o$ after ultralight EP reached $2.4 \times 10^{10}$, which was 1.5 times higher than the baseline $Q_o$ at the same field.

To better understand the nature of RF losses, residual and BCS components of the surface resistance ($R_s = G/Q_o$) were estimated for all three surface conditions. The Residual surface resistance ($R_{res}$) was extracted from the $Q_o$ values measured at a low temperature of ~1.5 K, while the BCS surface resistance ($R_{BCS}$) at 2 K was calculated by subtracting $R_{res}$ from $R_s$. Figure 5 shows a comparison of $R_{BCS}$ and $R_{res}$ for three surface conditions.

After mid-T baking, $R_{BCS}$ decreased significantly compared to the baseline, and $R_{res}$ also showed a reduction, consistent with earlier observations [3]. Ultralight EP further reduced $R_{res}$ with a small additional reduction in $R_{BCS}$.

The results suggest that the top RF layer of the mid-T baked cavity likely contained contaminants, which could be responsible for the observed poor performance. The application of minimal surface removal via ultralight EP appears effective in improving both $Q_o$ and $E_{acc}$.

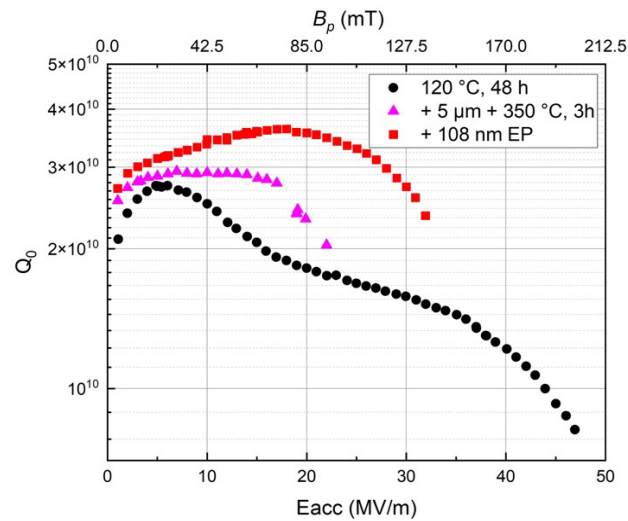

Figure 4: Q-E curves measured at 2 K after the 120°C bake, the mid-T bake at 350°C, and ultralight EP applied to mid-T baked surface.

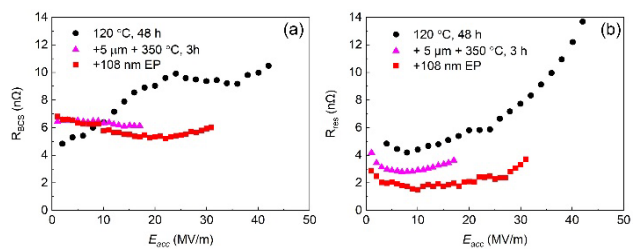

Figure 5: $R_{BCS}$ at 2 K (a) and $R_{res}$ (b) measured for three cases, including the 120°C bake, the mid-T bake at 350°C, and ultralight EP applied to mid-T baked surface.

Table 1: Cavity performance at 2 K after 120°C bake, mid-T bake at 350°C, and post-mid-T-bake ultralight EP.

| Pre-vertical test process | $Q_o$ at 17 MV/m | Quench field (MV/m) | $B_p$ at quench field (mT) |
|---|---|---|---|
| 120 °C, 48 h | $1.9 \times 10^{10}$ | 47 | 200 |
| Mid-T bake | $2.8 \times 10^{10}$ | 22 | 93.5 |
| Mid-T bake + Ultralight EP | $3.6 \times 10^{10}$ | 32 | 137 |

## SAMPLE STUDY

To gain deeper insight into the surface modifications responsible for the observed changes in RF performance, a parallel study was conducted using Nb samples. Two samples were subjected to mid-T baking at 350°C for 3 h in the same furnace used for the cavity. After baking, one of the samples underwent ultralight EP, removing ~120 nm of material.

Both samples were analyzed using Secondary Ion Mass Spectrometry (SIMS) to obtain elemental depth profiles. Figure 6 shows normalized intensities of oxygen (O), carbon (C), and niobium carbide (NbC), relative to Nb intensity as a function of sputtering depth.

The O profiles indicate that ultralight EP did not alter the O concentration, confirming the preservation of the O-enriched near-surface layer formed during the mid-T bake. A notable difference was observed in NbC signal, which was higher in the sample that did not receive ultralight EP. This elevated NbC intensity was observed within the top 20 nm of the surface. Removing ~100 nm in ultralight EP should obviously remove the surface carbides.

## DISCUSSION

Mid-T baking at 350 °C can lead to the formation of niobium carbides on the cavity surface. SIMS analysis of the Nb samples baked under identical conditions confirmed the presence of surface carbides, consistent with earlier reports [5]. The low quench field of 22 MV/m after mid-T baking is likely attributed to these carbides present within the RF layer.

While mid-T bake enhanced $Q_o$ of the cavity, anti-Q slope was not observed. This might be due to the presence of carbides, which enhanced $R_{res}$. The detrimental impact of these surface contaminants on RF performance was confirmed when ultralight EP, removing only 108 nm, significantly improved both $Q_o$ and quench field. Specifically, the quench field increased to 32 MV/m, and a clear anti-Q slope was observed, indicating improved performance primarily driven by a reduction in $R_{res}$.

Although adding ultralight EP introduces an additional step in the cavity processing, it can mitigate performance degradation, improve yields of high-performing cavities, and reduce variability across laboratories that operate furnaces under different vacuum and outgassing conditions (e.g., partial pressures of CO and $CO_2$), which might be responsible for the formation of carbides on the surface.

Since mid-T baking at 300°C yields a higher $Q_o$ compared to 350°C treated 1.3 GHz cavities [3], it would be interesting to extend this study to 300°C baked cavities.

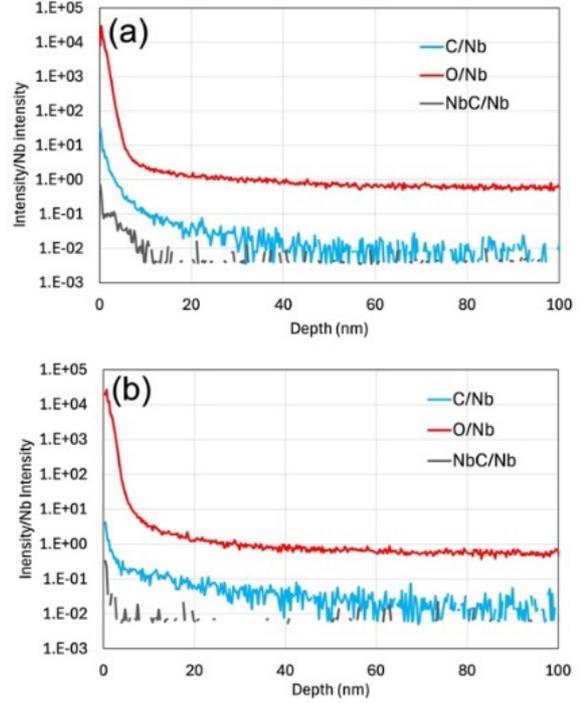

Figure 6: SIMS depth profiles of O, C, and NbC for (a) a mid-T baked Nb sample at 350°C for 3 h and (b) a similarly baked sample followed by 120 nm ultralight EP.

## CONCLUSION

Mid-T baking at 350°C for 3 h was applied to the 1.3 GHz single-cell cavity to improve its $Q_o$. While the treatment enhanced $Q_o$, the cavity quenched at a low field of 22 MV/m and showed no anti-Q slope in the corresponding Q-E curve. Application of ultralight EP (108 nm removal) significantly improved the cavity performance. The $Q_o$ value at 17 MV/m increased from $2.8 \times 10^{10}$ to $3.6 \times 10^{10}$, accompanied by a clear anti-Q slope. This ultralight removal also enhanced the quench field to 32 MV/m. These improvements suggest that the poor performance after mid-T baking may be attributed to the presence of NbC formed during the bake, as indicated by SIMS analysis of treated Nb samples. Further study is needed to reproduce these results and to explore whether similar benefits can be achieved for 300°C baked cavities.